
\newif\ifapj 
	\magnification=\magstep1
\else
	\magnification=\magstephalf
\fi
\input epsf
\epsfverbosetrue
\font\caps=cmcsc10
\font\bb=cmbx12
\tolerance=10000 \pretolerance=5000
\looseness=-3
\widowpenalty 10000
\displaywidowpenalty 10000
\overfullrule 0pt

\ifapj\baselineskip 24pt
 \else \baselineskip=15truept
\fi
\parindent=2.5em
\def\ub{\underbar}
\def\hi{\par\noindent \hangindent=2.5em}

\newcount\eqnumber
\eqnumber=1
\def\eqnum{\the\eqnumber\global\advance\eqnumber by 1}
\def\ls{\vskip 12.045pt}
\def\ni{\noindent}
\def\et{{\it et\thinspace al.}\ }
\def\eg{{\it e.~g.}\ }
\def\kms{km\thinspace s$^{-1}$ }
\def\deg{\ifmmode^\circ\else$^\circ$\fi}    

\def\msun{M_\odot}
\def\arcs{\ifmmode {'' }\else $'' $\fi}     
\def\arcm{\ifmmode {' }\else $' $\fi}     
\def\buildrel#1\over#2{\mathrel{\mathop{\null#2}\limits^{#1}}}
\def\mper{\ifmmode \buildrel m\over . \else $\buildrel m\over .$\fi}
\def\hper{\ifmmode \rlap.^{h}\else $\rlap{.}^h$\fi}
\def\sper{\ifmmode \rlap.^{s}\else $\rlap{.}^s$\fi}
\def\arcsper{\ifmmode \rlap.{' }\else $\rlap{.}' $\fi}
\def\arcmper{\ifmmode \rlap.{'' }\else $\rlap{.}'' $\fi}

\def\aj{{AJ}, }
\def\apj{{ApJ}, }
\def\apjs{{ApJSup}, }
\def\apjl{{ApJLett}, }

\def\mn{{MNRAS}, }

\def\spose#1{\hbox to 0pt{#1\hss}}
\def\lta{\mathrel{\spose{\lower 3pt\hbox{$\mathchar"218$}}
     \raise 2.0pt\hbox{$\mathchar"13C$}}}
\def\gta{\mathrel{\spose{\lower 3pt\hbox{$\mathchar"218$}}
     \raise 2.0pt\hbox{$\mathchar"13E$}}}

\def\hmpc{$h^{-1}$\thinspace Mpc}
\def\hkpc{$h^{-1}$\thinspace kpc}

\ifapj\vglue 29.10pt \fi

\centerline{\bb VELOCITY BIAS IN CLUSTERS}
\vskip 1.0truecm
\noindent
\centerline{\bf R. G. Carlberg}
\vskip 6pt
\noindent
\centerline{Department of Astronomy, University of Toronto,}
\centerline{60 St.~George St., Toronto, Ontario, M5S 1A1, Canada}

\bigskip\bigskip
\ifapj
\centerline{accepted: $\underline{\hbox to 6truecm{\hphantom\null}}$}
\vfill\eject
\fi

\def\bvs{\ifmmode{b_v(1)}\else$b_v(1)$\fi}
\def\bvp{\ifmmode{b_v(2)}\else$b_v(2)$\fi}

\centerline{ABSTRACT}

\ls
Hierarchical gravitational clustering creates galaxies that usually do
not fully share the dynamical history of an average particle in the
mass field.  In particular, galaxy tracers identified in numerical
simulations can have individual velocity dispersions in virialized
regions a factor \bvs\ lower than the dark matter. The field average
of the pairwise velocity dispersion depends on the statistical
weighting of collapsed regions, so that the tracer pairwise dispersion
is a different factor, \bvp, times the density field value.  A model
of a cool equilibrium tracer population demonstrates that mass to
light segregation is very sensitive to single particle velocity
bias. For the $\phi=-GM/(r+a)$ potential a $\bvs=0.9$ tracer indicates
a virial mass about a factor of 5 low.  The likely value of \bvs\ is
estimated and the simple equilibrium model is tested for applicability
using a $10^6$ particle simulation of the formation of a single
cluster from cosmological initial conditions.  A striking outcome is
that the dense central cores of infalling galaxy scale dark matter
halos survive infall and virialization within the cluster.  These
dense cores do not have the dissipationless infall of the bulk of
mass, but during their first cluster crossing lose energy to the
cluster background to become systematically more bound, thereafter
orbiting freely as coherent, self-gravitating units.  From this
simulation, the value of
\bvs\ is estimated as $0.8\pm0.1$.  The pairwise velocity dispersion
bias, \bvp, which is equal to \bvs\ augmented with any
anti-bias of galaxies against high velocity dispersion clusters.  The
value of \bvp\ is estimated to be 15\% less than \bvs\ on the basis of
a biasing model where the M/L ratio rises a factor of 10 from binaries
to clusters, consequently $\bvp=0.6^{+0.2}_{-0.1}$.  The Cosmic Virial
Theorem measures the combination $b^2_v(2)\Omega/b$ (where b is the
clustering bias) to be in the range 0.20-0.36, which at $\bvp=0.6$ is
compatible with $\Omega=1$ for clustering bias $b\simeq1$ (IRAS
galaxies) to $b\simeq1.8$ (optical galaxies) but is incompatible with
the full COBE normalized CDM spectrum.  Comparison of cluster mass and
luminosity profiles at large radii is a test for the existence of
single particle velocity bias.

\vskip 40pt

\vfill\eject

\ni\ub{1. INTRODUCTION}

\ls
Although there are compelling theoretical arguments favoring an
$\Omega=1$ universe (\eg Guth 1981, Bardeen \et 1983) a substantial
body of observational data indicates $\Omega\simeq0.2$.  That is, if
$\Omega=1$ then the pattern of galaxy cluster should lead to random
pairwise velocities of galaxies of 600 to 1000 \kms.  The observed
random pairwise velocities are 2 to 3 times smaller than the
$\Omega=1$ prediction (depending on galaxy population, Davis and
Peebles 1983, Bean \et 1983, Davis \et 1985, Hale-Sutton
\et 1989, Fisher \et 1993). The
virial theorem applied to rich clusters of galaxies also finds
$\Omega\simeq0.2$ (\eg Kent and Gunn 1982).  On the other hand,
significantly larger $\Omega$ values are found in studies using
velocity data on scales where the clustering is essentially
linear. The combination $\Omega^{0.6}/b$ ($=\sigma_8\Omega^{0.6}$,
where $b=1/\sigma_8$ is the bias, the ratio of galaxy number variance
to total density variance, $\sigma_8$, in 8\hmpc\ spheres) is estimated
to 0.4 (Lynden-Bell \et 1989) 0.86 (Kaiser \et 1991), 0.6 (Strauss \et
1992), and 1.3 (Nusser and Dekel 1993), all with fairly wide
confidence intervals.  To varying degrees the large scale flow
measurements are all incompatible with the Cosmic Virial Theorem
(Peebles 1976) $\Omega$ estimates, $\Omega/b\simeq0.2$ (Davis \&
Peebles 1983) and $\Omega/b=0.38$ (Fisher \et 1993).  The important
conclusion is that either $\Omega\neq1$ (for a single component dark
matter), or, galaxies must have a clustering amplitude greater than
the density field (Kaiser 1984), or their velocity dispersions must be
reduced below the density field, or some combination of these. The
possibility that the random velocities are affected yields a scale
dependent dynamical bias which can reconcile the low $\Omega$ values
in clusters with the $\Omega\simeq1$ from large scale flows.

A velocity bias, in which the galaxies have smaller {\it pairwise}
random velocities than the dark matter,
$\bvp=\sigma_{12}(\hbox{galaxies})/\sigma_{12}(\rho)\simeq 0.5-0.8$
has been reported in a number of cosmological simulations that have
hierarchical galaxy formation (Carlberg and Couchman 1989, Carlberg,
Couchman and Thomas 1990, Couchman and Carlberg 1992, Cen and Ostriker
1992, Gelb and Bertschinger 1993) but was not detected in other
physically similar simulations (Katz \et 1992). The effect exists only
in the strongly nonlinear regime---infall velocities are close to
their expected values (Carlberg 1991).  The single galaxy velocity
bias in virialized clusters, which excludes statistical effects of
pairs, has been found to be
$b_v(1)=\sigma(\hbox{galaxies})/\sigma(\rho)\simeq0.7-0.9$ (Carlberg
and Dubinski 1991, Evrard \et 1992, Katz and White 1993). The single
particle velocity bias, \bvs, is the most readily testable aspect of
velocity bias, and can be constrained from observational studies of
clusters which examine the mass profile of the cluster beyond the
apparent virial radius.

The quantitative significance of velocity bias is largely an issue of
how galaxies are formed.  Cosmological gasdynamical simulations of
galaxy formation is a vigorous research field, but the star formation
process remains poorly understood.  The approach here is to
concentrate on the dynamical problem, which is simplified in
dissipationless experiments, and is a valid approximation for the
available resolution since the gravitational field is everywhere
dominated by background dark matter on all scales.  The presence of
velocity bias in a cluster is equivalent to an energy loss of the
galaxy tracer population. Since the origin of the effect is
controversial, an aspect of this paper is to show that the effect
arises in the absence of a viscous or dissipating gas component.  This
is not to say that gasdynamics doesn't have an crucial role to
play---it is the essence of the galaxy formation process. It should of
course be noted that several gasdynamical simulations have found a
single particle velocity bias in simulated clusters (Evrard \et 1992,
Katz \& White 1993).  The aim of this paper is to show that velocity
bias has a dynamical origin which will hold for a broad class of
galaxy formation scenarios.

The approach to galaxy identification here is that proto-galaxies are
assumed to form at $z\simeq1-3$ in the dense ($\rho\gta10^5\rho_0$)
central regions of dark matter halos having velocity dispersions in
the range around 100 \kms (White and Rees 1978), and then, once
formed, these proto-galaxies may merge with others to build up the
$z=0$ galaxies.  This paper reports the results of a simulation having
sufficient particles and length resolution that the dense cores of
halos contain the same particles over a Hubble time, thereby
diminishing the parameter dependence of the galaxy selection
algorithm. The long lived cores that are visible can easily be picked
out using a variety of criteria.  With this relatively straightforward
approach for tracing a common set of particles that remain in close
proximity in a dense, self-gravitating region (like a galaxy) a single
object velocity bias clearly emerges along with a readily measured
relative energy loss between core and tidally removed halo. The
orbital kinematics of these galaxy tracers, and the differences in
their dynamics from their loosely bound outer halos are examined.

Observational tests for the existence and magnitude of velocity bias
are best done on clusters of galaxies. The goals of this paper
are to predict the expected velocity bias, and show its relationship
to the relative density distributions of the galaxies and the cluster
mass.  The next section uses a realistic cluster potential to show
that a relatively small velocity bias gives rise to a relatively large
``mass segregation'' (between the slightly cooled galaxy population
and the cluster mass). The short \S3 models the likely reduction of
\bvp\ from \bvs. To predict \bvs\ the kinematics of various galaxy
tracers in a large simulation of the formation of a cluster
of galaxies are compared to the kinematics of the mass field in \S4.
Section 5 demonstrates that the particles identified as galaxy
tracers systematically lose energy in comparison to average particles
over the course of their infall into the cluster. Section
6 relates velocity bias to the interpretation of cluster masses and
the field density, and states the results of this paper.

\ls\ls\goodbreak
\ni\ub{\bf 2. VELOCITY BIAS AND MASS SEGREGATION}

\ls
A cool, tracer population in an equilibrium system with a power law
mass density profile, $\rho(r)=b_\rho r^{-n}$, is in general more
centrally concentrated than the mass distribution.  For a spherical
system with an isotropic velocity ellipsoid the self-gravitating
``background'' mass has a velocity dispersion
given by the Jeans equation, $\nabla(\rho\sigma^2) = -\rho\nabla\phi$,
to be, $$
\sigma^2_b(r)\equiv b_\sigma r^{2-n} =
{{4\pi G b_\rho} \over {2(n-1)(3-n)}} r^{2-n}.
\eqno{(\eqnum)}
$$ Equation (1) is valid over the range $1<n<3$, which is the local
density gradient found for all but the inner and outermost part of a
typical dark halo (Dubinski and Carlberg 1991, Hernquist 1992).  A
cool tracer population with velocity dispersion
$\sigma^2_c(r)=c_\sigma r^{-q}$ and density profile $\rho_c(r) =
c_\rho r^{-p}$ inserted into this background potential must
in equilibrium satisfy
the Jeans equation,
$$ (p+q) c_\sigma r^{-q-1} = 2 (n-1) b_\sigma
r^{-n+1},
\eqno{(\eqnum)}
$$
from which $q=n-2$, the same radial dependence
of the velocity dispersion as for the background population.
Noting that $c_\sigma/b_\sigma=b^2_v(1)$,
$$
p = n + 2(n-1) \left( {1\over{b^2_v(1)}} -1\right).
\eqno{(\eqnum)}
$$
Hence a cool tracer population, $b_v(1)<1$, will have $p>n$. That
is, the tracer will be a steeper, more centrally concentrated
distribution.  Because the exponent of the power law depends on the
ratio of velocity dispersions, a small temperature difference
translates into a large radial segregation.  The background mass
interior to radius $r$ is $M(r)=4\pi b_\rho r^{3-n}/(3-n)$, with a
similar relation for the cool population. If both masses are
normalized at some common outer radius, half the turnaround radius for
instance, then the ratio of the half mass radii is,
$$ {r_{1/2}^c
\over r_{1/2}^b} = \left({1\over
2}\right)^{{1\over{3-p}}-{1\over{3-n}}}.
\eqno{(\eqnum)}
$$
If the background profile is taken as ``isothermal'', $n=2$, and
the cool population has a velocity dispersion
$\sigma_c=\sqrt{4/5}\sigma_b=0.894\sigma_b$, giving $p=5/2$, then
$r_{1/2}^c/r_{1/2}^b=1/2$, a dramatic effect.

\setbox17=\vbox{
\noindent \ifapj\else\narrower\baselineskip 12pt\fi
Figure 1:\hskip 5mm The equilibrium density profile for a cooled
population in a CDM like dark halo. The $\rho(r)$ are given as a
function of decreasing \bvs. The top curve is for $b_v(1)=1$, with
decrements of 0.05 to the bottom, where $b_v(1)=0.8$.  In all cases
the half mass radius is scaled to be one unit, and the density there
is scaled to unity. Note that only the outer part of the profile varies
substantially with \bvs.  Approximate fits are given in the text.}
\ifapj\else
\midinsert
\epsfysize=3.0truein
\centerline{\epsfbox[24 200 588 688]{den.ps}}
\copy17 \endinsert \fi

A more realistic model of the effect of a cooled population is to use
the Hernquist (1990) potential $\phi(r) = -GM/(r+a)$ as the
background, which has been found to be a good fit to the halos found
in CDM model universes (Dubinski and Carlberg 1991) and merged
galaxies (Hernquist 1992).  A limitation of the assumption of
equilibrium is that it does not account for infall which is important
at large radii (typically twice
the nominal half mass radius).  The potential will be taken to be spherically
symmetric, and the tracer population to have an isotropic velocity
dispersion. In this case Jeans equation for stellar hydrostatic
equilibrium with an isotropic velocity dispersion integrates to give
the density profile of the tracer population,
\newcount\eqjeans \eqjeans=\eqnumber
$$
\ln{{\rho(r)}\over{\rho(a)}} =
-\ln{{\sigma^2(r)}\over{\sigma^2(a)}}
-\int_a^r {1\over{\sigma^2(x)}} {{d\phi(x)}\over{d x}} \, dx ,
\eqno{(\eqnum)}
$$ where $a$ is the scale radius, and $\sigma(r) =
b_v(1)\sigma_{\rho}$ and $\sigma_{\rho}$ is the self-consistent
solution for the Jeans equation (Hernquist 1990) $$
\sigma_{\rho}^2(r) = {{GM}\over{12a}}
	\left\{ {{12r(r+a)^3}\over{a^4}}
         \ln{\left({{r+a}\over r}\right)
   -{r\over{r+a}}}\left[ 25+52{r\over a} +42\left({r\over a}\right)^2
   +12\left({r\over a}\right)^3 \right] \right\}.
\eqno{(\eqnum)}
$$

Solutions to Equation (\the\eqjeans) are displayed in Figure~1 as
a function of \bvs.  The
density profiles of Figure~1 are not expected to be a good fit to
clusters in the innermost regions, since galaxy merging and stripping
will reduce their numbers. The large reduction in scale radius
is a necessary consequence of the energy loss of
the infalling galaxies (discussed in section 4) which leads to a
slight drop in velocity dispersion of the tracer population.  Near the
center, the number density of galaxies is likely diminished below the
model shown here due to the cluster tidal field and galaxy merging.  A
fit, good to about 50\%, to the tracer density profiles shown in
Figure~1 is,
\newcount\eqrhofit \eqrhofit =\eqnumber
$$
n_t(r)\simeq {{(p-2)(p-1)}\over{r(1+r)^p}},
\eqno{(\eqnum)}
$$
where the profile shape, $p$, and the half mass radii, $r_{1/2}$ are
approximately,
$$
p\simeq 4\tan{[0.6435+1.8435(1-b_v(1))]},
	\quad r_{1/2}\simeq2.414\times 10^{-4.012(1-b_v(1))}.
\eqno{(\eqnum)}
$$
The tracer luminosity
contained in $r$ is the normalized volume integral of
Equation (\the\eqrhofit),
$$
L_t(r) = 1 - {{r^2(p-1)+rp+1}\over{(1+r)^p}}.
\eqno{(\eqnum)}
$$
The virial radius as indicated by the
tracer population is even more sensitive to \bvs\ since
$
r_{v}^{-1}(L) = L^{-2}\int_0^\infty 4\pi L_t(r) n_t(r) r dr
$
involves two integrals over the tracer mass profile.
An approximate fit to the virial radius (in units where
the scale radius of the potential is one unit) is,
$$
r_v = 6.0\times10^{[-8.78+8.42(1-\bvs)](1-\bvs)}.
\eqno{(\eqnum)}
$$

\setbox21=\vbox{
\noindent \ifapj\else\narrower\baselineskip 12pt\fi
Figure 2:\hskip 5mm The density-radius relation (left, normalized
to $\rho_0$ assuming a background expanding as $t^{2/3}$) and the velocity
dispersion radius relation (right) for all the mass (solid line),
particles linked with $0.2\overline{d}$ at z=2 (asterisks), particles
linked with $0.2\overline{d}$ at z=2 and $0.02\overline{d}$ at z=0
(circles), and the bound ``galaxies'' which the circle particles
make. These ``galaxies'' are eroded in the center. Note that low
velocity populations have steeper density profiles. The particles
identified with circles have $\bvs=0.84$, and indicate a virial radius
and mass that are, respectively, 0.22 and 0.15 of the density
field values.}
\ifapj\else
\midinsert
\epsfysize=3.0truein
\centerline{\epsfbox[24 476 588 740]{one.ps}}
\copy21 \endinsert \fi

The comparison of the model to the cluster data of Section 4 displayed
in Figure~2 shows that the above model works well in the radial range
the cluster is virialized, which is about twice the half mass radius
of the cluster. Beyond that radius all components are fairly well
mixed, as is expected for infall. The simplest components are the
individual cool particles which are well described by the model. The
coolest component has $\bvs=0.84$, a virial radius from the tracer of
0.33\hmpc\ (22\% of the true value), and a virial mass which is 15\%
of the true value. These results are in reasonable accord with the
equilibrium values predicted above.

\ls\ls
\goodbreak
\ni\ub{\bf 3. VELOCITY BIAS IN THE FIELD}

\ls
The pairwise velocity bias in the field population,
\bvp, is \bvs\ averaged over virialized regions augmented
with a statistical bias if high velocity dispersion clusters have
fewer galaxies than low velocity dispersion regions (Bertschinger and
Gelb 1991, Couchman and Carlberg 1992, Gelb and Bertschinger 1993).
The statistical pairwise velocity reduction is likely
about 10\%, as is estimated below assuming most pairs closer
than $\sim$1\hmpc\ are in the same virialized halo. The numbers
of halos in a given mass range is described by the
Press-Schechter (1974) distribution of halo masses, $n(M)d\ln{M}
\propto \rho_0M^{-1} \exp{[-\nu^2/2]} \nu\gamma(M)\, d\ln{M}$, where
$\gamma$ is a very weak function of $M$ that will be assumed constant.
The number of galaxy pairs in halos of mass $M$ is proportional to
$M^2n(M)d\ln{M}$. The peak height parameter, $\nu\propto M^{(n+3)/6}$,
for a power law density perturbation power spectrum, $P(k)\propto
k^n$. Clustering observations indicate that $n\simeq -1$ in the
cluster mass range and therefore $\nu\propto M^{1/3}$.  Therefore the
RMS pairwise velocity dispersion created by pairs of galaxies in
virialized units is,
$$
\langle \sigma_p^2 \rangle =
{
{\int_{\nu_c}^\infty b^2_p(\nu)\sigma^2(\nu)
e^{-\nu^2/2} \nu^3 \, d\nu }
 \over
{\int_{\nu_c}^\infty  b^2_p(\nu)
e^{-\nu^2/2} \nu^3 \, d\nu }
},
\eqno{(\eqnum)}
$$ where $\nu M d\ln{M} \propto \nu^3 d\nu$.  The function $b_p(\nu)$
allows for a possible change of numbers of galaxies per unit cluster
mass in clusters of increasing velocity
dispersion.  A simple linear anti-clustering model is,
\newcount\eqbias \eqbias=\eqnumber
$$
b_p(\nu)={\nu_0\over\nu} (1-b_\infty)+b_\infty,
\eqno{(\eqnum)}
$$ which has $b_p=1$ at $\nu=\nu_0$, and declines to $b_p=b_\infty$
for large $\nu$, as is suggested by an increasing M/L with increasing
cluster velocity dispersion.
The proportionality constant is the $\nu_0$ associated
with the smallest groups of galaxies.

A practical set of parameters is $\nu_0\simeq 0.5$ which is
appropriate for 1.5 \hmpc\ tophat in unbiased CDM model. The greatest
possible bias is about $b_\infty=0.1$ as is suggested by an apparent
M/L increase from 30$h$ in binaries (White \et 1983) to about 300$h$
in clusters (\eg Kent and Gunn 1982, David and Blumenthal 1992).
Virialized groups found in simulations (Couchman and Carlberg 1992)
have precisely this sort of rise in their total mass per ``galaxy''
contained. The model then predicts that the statistical effect is a
15\% reduction in the velocity dispersion.  Larger $\nu_0$ and
$b_\infty$ closer to unity both decrease the size of the statistical
bias, so 15\% is close to the upper limit.  This statistical
reduction combined with the $\bvs=0.8\pm0.1$ found below gives
$\bvp\simeq0.6^{+0.2}_{-0.1}$.  Note that if there is any statistical
velocity bias it is essential that $\bvs<1$, otherwise clusters would
have M/L values exceeding the closure M/L.  The model of
Equation~(\the\eqbias) and the parameters chosen are consistent with
M/L data, but can also be regarded as the upper limit to the reduction
of \bvp\ below \bvs.

\ls\ls
\goodbreak
\ni\ub{\bf 4. DISSIPATIONLESS CLUSTER FORMATION}

\ls
The formation of large cluster like Coma has been the subject of many
numerical simulations. The primary reason for repeating what might
appear to be the same experiment with ever more computer power is that
many of the fundamental applications of clusters dynamics to
cosmological parameter estimation remain uncertain.  A basic problem
is that visible galaxy masses are less than $10^{-4}$ of the cluster
mass, meaning that until $N\sim 10^4$ in a single cluster, individual
particles are more massive than single luminous galaxies. With $N\gg
10^4$ particles a dynamical definition of the site of a galaxy using
groups of particles becomes possible (\eg White \et 1987, Carlberg and
Couchman 1989, Couchman and Carlberg 1992). The understanding of
initial conditions and boundary conditions has changed substantially
from the early simulations, for instance tophat collapses with no
shear fields, while dynamically informative, are not useful for the
details of the dynamics of cluster collapse in currently favored
cosmologies (where shear fields are important and the cluster buildup
occurs over the age of the universe).

Two body relaxation remains a problem even when $N\sim10^6$ and the
self-relaxation time in a cluster is of order $10^3$ crossing times.
That is, galaxy sites are low internal velocity dispersion, cool,
halos orbiting in the hot background particles of the cluster which
can heat the galaxy halo until it boils away.  A quantitative estimate
of the location in the cluster where evaporative destruction becomes
significant is straightforward.  At the half mass radius of a cluster
containing $N$ particles at velocity dispersion $\sigma_c$ the ratio
heating timescale of a cool halo with velocity dispersion
$\sigma_g<<\sigma_c$ to the crossing time is approximately, $$
{{t_{heat}(g) }\over t_c} = {N \over{16\ln{[r_{1/2}/\epsilon}] }}
\left({\sigma_c\over\sigma_g}\right)^2
\eqno{(\eqnum)}
$$ (Spitzer 1962, Binney and Tremaine 1987, Huang \et 1993).
Parameters relevant to the simulation here are
$\sigma_g\simeq 100$ \kms and $\sigma_c\simeq 1500$ \kms (both one
dimensional) and $N=10^6$, for which the heating time is about 10
crossing times at the half mass radius, and one crossing time at
$0.4r_{1/2}$, which is 0.45\hmpc\ for the cluster studied here.  For
the resolution available, about 10 kpc, the absence of a dissipated
``stellar'' component at the center of a
galaxy halo makes no significant difference to
the outcome.  Increasing the softening does not help, since the
heating time scales as $\ln{(r_{1/2}/\epsilon)}$, where
$\epsilon\gg 2r_{1/2}(\sigma_c/\sigma_g)^2/N$ to be in the regime of
small angle deflection.  For instance, in the simulation
$\epsilon\simeq10^{-2}r_{1/2}$ so that a doubling of $\epsilon$ gives
about a 1\% increase of the two-body heating time scale. The
considerable disadvantage of doubling the softening is that the
self-energy of the smallest substructures is reduced by a factor of
two, making them {\it easier} to evaporate.

The heating rate derived from hot-cold energy transfers is an upper
limit to the dissolution time (which, however, appears to describe the
dissolution of substructure in the simulation done here reasonably
well). In detail substructure dissolution remains a research problem.
The dissolution of small bound units in a tidal field has been studied
in the context of globular clusters (\eg Chernoff \& Weinberg
1990). The additional features here, softened gravity (so that no
binaries can ever be ``hard'' which is the dominant
energy source for evaporation), a hot background of equal mass
particles, and infall with an accompanying halo substantially reduces
the direct applicability of their results.

\setbox11=\vbox{
\vfill
\noindent
Figure 3:\hskip 5mm The xy projection of the particles in a box of
edge 4 \hmpc\ around the cluster center, at a scaled age of 15.8~Gyr.
Most of the visible dense substructures are the cores of infalling
substructures and have orbited without further tidal truncation or
dynamical friction, for at least one crossing time.  }
\ifapj\else \pageinsert \copy11
\medskip
\epsfysize=7.6truein
\centerline{\epsffile{pdm.384.ps}}
\endinsert \fi

\setbox12=\vbox{
\vfill
\noindent
\ifapj\else\baselineskip 10pt\fi
Figure 4:\hskip 5mm Phase space variables in the central region of the
cluster (a 2 \hmpc\ box around the cluster center) at the same time as
Figure~3, 15.8~Gyr. The upper part of the figure shows the logarithm
of the velocity dispersion, and the lower part the square root of the
volume density, both functions chosen simply to compress the range of
the data. The ``haze'' at the top is cluster particles at a velocity
dispersion of 1000 \kms. The ``rain'' is substructure with internal
velocity dispersion of 100 to 300 \kms.  The xy plane has been tilted
by 1.5\deg\ to provide some separation of physically removed regions.
The substructures that visible in the xy plot of Figure~3 are the
dense regions at low velocity dispersion. Note that the macroscopic
phase space density effectively has a discontinuity between the hot
cluster population and the cool pieces of substructure which orbit
independently within the cluster. The velocity and density dispersion
are calculated using each particle's 8 nearest neighbors.  }

\ifapj\else \pageinsert \copy12
\medskip
\epsfysize=7.0truein
\centerline{\epsfbox[22 30 588 740]{pdsc.384.ps}}
\endinsert \fi

\goodbreak\ls\ni\ub{\it 4.1 Numerical Details}

The initial conditions for the simulation were created using the CDM
spectrum (Bond and Efstathiou 1984), realized at an initial amplitude
of $\sigma_8=0.1$ on a $128^3$ grid in a box 50\hmpc\ on a side. The
position of the largest peak in the box was approximately located by
smoothing with a 2.5\hmpc\ Gaussian filter, and adjusted by doing a
few $N=3\times10^4$ trial runs.  A sphere of 12.5\hmpc\ co-moving
radius around the peak was loaded with $10^6$ randomly placed
particles (each of mass $2.27\times10^9h^{-1}\msun$), perturbed with
the Zeldovich approximation and extracted for evolution with a
multistep, quadrupole treecode (Barnes and Hut 1986, Dubinski 1988).
The internal gravitational field was supplemented with an external
quadrupole tidal field with an amplitude evolving as the linear theory
variance of the tidal field on a 12.5\hmpc\ sphere.  Some attention to
boundary conditions is essential, since a vacuum boundary creates
orbits that are far too radial in the outer regions of the cluster.
The pairwise Plummer softening is $\epsilon=9.7$ \hkpc, and the
opening angle parameter was set at $\theta_c=0.8$.  The minimum time
step is set at $\Delta t=10$~Myr with the final time being 13.0~Gyr
(1300 steps). The 13~Gyr epoch formally corresponds to a $\sigma_8=1$
linear normalization.  The time displayed in Figures~3 and 4 formally
is 15.8~Gyr to allow the cluster to relax a bit after a major
merger. The precise correspondence for any times over 13~Gyr is
weakened by the fact that the outer boundaries begin to fall into the
cluster.  The simulation took approximately 3 cpu-months on a 4 Mflop
SGI machine.

The xy projection of the final state of the million particle cluster
simulation shown in Figures~3 and 4 has the important feature of a
substantial amount of low mass substructure that has survived a number
of orbits within this well virialized cluster. The innermost region
has a deficiency of substructure as a result of two body evaporative
heating. That is, after the objects are identified the test as to
whether their internal velocity dispersion is too large to be
self-bound selectively rejects objects near the center of the cluster.
The final cluster is a triaxial anisotropic object, containing a
relaxed mass of $M=2.2\times10^{15} h^{-1}\msun$. The spherically
averaged half mass radius, $r_{1/2}$, is 1.13 \hmpc, and the RMS
velocity dispersion is 1080 \kms (one dimensional).  The velocity
dispersion is nearly isotropic in the center, has a peak value of 1530
\kms\ at 0.3\hmpc, and beyond that becomes mildly radial, with the
tangential components each being about 2/3 of the radial component.
The density profile is well fit with $\rho\propto r^{-1}(r+a)^{-3}$
with $a=0.47$\hmpc\ (see Figure~2). As far as can be measured the
central density diverges approximately as $r^{-1}$, with no core
---the only physical length in this cold collapse is the turnaround
radius of the cluster, so there is no physics to set a core radius
(Dubinski and Carlberg 1991).

\ls\ni\ub{\it 4.2 Galaxy Tracers in the Cluster}

The dense pieces of substructure visible in Figure~3 are durable
self-gravitating regions with a velocity dispersion around 100 \kms\
and clump masses of order $10^{-4}$ of the cluster mass. Their
mass-velocity dispersion relation implies that most of the
substructures are typically a factor of 10 light compared to the
normal relation, a result of tidal stripping removing their low
density outer parts.  Figure~4 shows a combined plot of velocity
dispersion and density which illustrates that the visible
substructures are completely distinct regions of high phase space
density. At the time displayed there is essentially no substructure
with a mass greater than $10^{-3}$ of the cluster mass, at least
partly due to merging and tidal truncation (Fall and Rees
1977).

Identifying the sites of galaxies in a simulation is a fundamental
problem for the application of simulation results to the observable
universe.  Early cosmological simulations assumed that the association
probability between galaxies and particles was uniform for all
particles (\eg Gott \et 1979). Painting particles at ``peaks'' of the
initial density field showed that particles at particular initial
sites could have substantially higher clustering amplitudes than the
mass field, but the velocities were not significantly different (Davis
\et 1985).  Galaxy formation within cosmological gas dynamical
simulations remains subject to uncertainty over the dominant physical
processes.  Within the dissipationless assumption, galaxies can be
identified as single particles or groups of particles can be merged
into a single heavy particle (White \et 1987, Summers 1993, van Kampen
1993). The single particle galaxy approximation requires additional
criteria to accommodate galaxy merging and any subsequent tidal
stripping, which are dynamically important, especially if the single
particle galaxy's mass has some dark matter background included.  The
issue of interest here is essentially dynamical--- how does mass
segregation evolve once the galaxies are formed?--- which can be most
clearly addressed in a dissipationless simulation.

In large N simulations there is a straightforward galaxy
identification procedure: locate groups of particles in the dense
central regions of halos having velocity dispersions in the range
around 100 \kms (Rees and Ostriker 1977, White and Rees 1978,
Blumenthal \et 1984, White \et 1987). Ultimately galaxy identification
will improve with gas dynamical simulation, however in the situation
that all the computing power available is only capable of marginally
resolving galaxy scales in a cluster simulation, with significant
relaxation problems remaining, the approach of this paper is to do an
entirely gravitational simulation, and to demonstrate that the dense
dark matter cores persist in the virialized cluster.  The existence of
clearly defined long lived substructures with the approximate velocity
dispersions of galaxy halos greatly relieves the problem of galaxy
identification---although it does not vanish completely.  The densest
parts of these objects can be reliably identified at $z\simeq5$ and
traced to the final epoch, accurately accounting for merging along the
way through the gravitational dynamics of the simulation.  Most of the
halos orbit for more than a crossing time, and therefore reliably
sample the potential of the cluster.

In detail the galaxy tracers identified here are selected from the
simulation as follows. At time 2.5~Gyr (z=2) all particles closer than
$0.2$ times the mean interparticle separation, $\overline{d}(t)$, are
linked into groups. This selects particles at a minimum overdensity of
approximately $125\rho_0$ which should be in collapsed structures
which will not easily unbind.  Groups with 10 or more particles,
$2.3\times10^{10}h^{-1}\msun$, are assumed to be ``proto-galaxies''.
At least two dynamical effects must be taken into account, tidal
shredding in the cluster and merging to create single
galaxies. Therefore, the selected ``proto-galaxy'' particles
naturally follow the galaxies orbit,
and can merge and be tidally distended. The tightly bound survivors are
relinked at 13~Gyr using a percolation length $0.02\overline{d}(t)$ to
pick out the dense cores and discard stripped off halo particles. The
final galaxy tracers are required to have 10 or more particles. Groups
that have an internal velocity dispersion that is greater than
450$(M/10^{12}M_\odot)^{1/3}$~\kms\ are discarded as being unbound or
severely contaminated with transient cluster particles. The result is
a list of about 100 tracer galaxies.  The average tracer contains 50
particles at a mean overdensity of $\gta10^5\rho_0$.  The dynamical
properties of these tracers are insensitive to factor of two changes
in the selection parameters.

The individual particles can be viewed as the most conservative
statistical tracers of galaxies. That is, the small swarm of
individual particles in the dense region of a halo can be numerically
grouped together to make a ``galaxy'', or each one can be viewed as an
equally valid galaxy. The main dynamical difference is that the
individual particles are orbiting in the halo, rather than being
nearly at rest in its rest frame. These particles exhibit all
the velocity bias effects. Grouping the particles together into
``galaxies'' allows the dynamics of merging and tidal stripping to be
incorporated, but has the drawback that N-body heating and evaporation
in the dense central region of the cluster systematically destroys the
groups that orbit in the center of the cluster.

\ls\ni
\ub{\it 4.3 Dynamical Mass Underestimates}

\setbox13=\vbox{
\noindent \ifapj\else\narrower\baselineskip 12pt\fi
Figure 5:\hskip 5mm Virial mass estimates normalized to the dark
matter virial mass at 13~Gyr.  The symbols mark the following
estimators (see text): plus--random sample, asterisk--(particles
$\le0.2\overline{d}$ at z=2) circle--(all particles
$\le0.2\overline{d}$ at z=2, and $\le0.02\overline{d}$ at z=0)
cross--(circle particles grouped into ``galaxies'', excluding cD),
square--(luminosity weighted ``galaxies'', including central ``cD'').
The spikes at times 9.4 and 13~Gyr occur during major cluster mergers.
}

\ifapj\else
\midinsert
\epsfysize=3.0truein
\centerline{\epsfbox[24 200 588 688]{vm.ps}}
\copy13 \endinsert \fi

A compendium of virial mass estimators of the cluster as a function of
time is exhibited in Figure~5. The masses are normalized to the
(constant) cluster mass at 13~Gyr as measured by counting the
particles in the virialized cluster. The uppermost line (with plus
signs) is the virial mass estimate based on a random subset of all the
particles in the cluster, and gives an accurate measure of the true
mass.  At early times the cluster is in nearly pure Hubble expansion
and the cluster is not in equilibrium.  The maxima in the mass
estimates at 9.4 and 13~Gyr are the result of mergers of the main
cluster and large infalling groups.  The other lines in Figure~5 give
mass estimators from low density galaxy halo particles, high density
halo particles (neither of which remove the internal motions of the
substructures), and the galaxy tracers, with and without the large
central galaxy.  Figure~6 displays the time varying ratio of the
virial mass estimates to the dark matter virial mass estimate,
emphasizing that any tracer located in structures that form at
moderate redshifts will underestimate the mass of the cluster.  No
tracer identified here is entirely satisfactory, due to the problems
of two body relaxation which evaporate low velocity dispersion
substructures (see Figure~8). At most 10\% of the object list is
discarded because the objects are unbound. The ``missing'' objects are near
the radius where the velocity dispersion is a maximum,  meaning that the
\bvs\ measured is slightly increased (likely about 5\%) over the
``true'' value.  In spite of this, the simulation is entirely adequate
to demonstrate the existence and approximate magnitude of velocity
bias.  Velocity bias can be measured in cluster simulations as small
as 4000 particles (Carlberg and Dubinski 1991), although no
substructure is retained in that case.

\setbox14=\vbox{
\noindent\ifapj\else\narrower\baselineskip 12pt\fi
Figure 6:\hskip 5mm Ratio of virial mass estimate to the dark matter
virial mass estimate at the same time.  The symbols mark the same
populations as in Figure~5.  The crosses and squares are the galaxy
tracers.  }

\ifapj\else
\midinsert
\epsfysize=3.0truein
\centerline{\epsfbox[24 200 588 688]{vmb.ps}}
\copy14 \endinsert \fi

The primary conclusion of Figures~5 and 6 is that the cluster mass is
always underestimated by any tracer which originated in the collapsed
substructures present prior to the formation of the cluster.  Tracers
that include low density halo material that is easily stripped off in
the cluster give a mass $\simeq$60\% of the full mass. Particles in
the high density cores give masses 15-40\% of the full values. The
``galaxies'' identified are subject to erosion via two-body heating in
the central regions, and as the objects in the central regions are
evaporated (see Figure~8) the population acquires a central density
dip that overestimates the mass in comparison to an artificial robust
population (see also van Kampen 1993).

\setbox15=\vbox{
\noindent\ifapj\else\narrower\baselineskip 12pt\fi
Figure 7:\hskip 5mm Single particle velocity bias, defined as the
ratio of the tracer population velocity dispersion to the mass field
velocity dispersion.  The symbols mark the same populations as in
Figure~5.  The average value of $b_v(1)\simeq0.85$ for the
``galaxies'' (cross and square markers) after the last cluster merger,
with this value likely being a slight overestimate of \bvs\ because
of the missing galaxies in the inner 200\hkpc.
}

\ifapj\else
\midinsert
\epsfysize=3.0truein
\centerline{\epsfbox[24 200 588 688]{vvb.ps}}
\copy15 \endinsert \fi

\setbox16=\vbox{
\noindent\ifapj\else\narrower\baselineskip 12pt\fi
Figure 8:\hskip 5mm Virial radii (solid lines) and half mass radii
(dashed lines) normalized to the time
varying dark matter radii for the same
populations as in Figure~5.  The rise in the half mass radii of the
galaxies (and their underlying ``stellar'' population) is a result of
central merging and two-body evaporation of the galaxy tracers.  The
virial radii indicate that the tracers form with a small radial bias,
which increases during cluster infall, remaining fairly steady once
the cluster has substantially virialized.  }

\ifapj\else
\midinsert
\epsfysize=3.0truein
\centerline{\epsfbox[24 200 588 688]{vrb.ps}}
\copy16 \endinsert \fi

Figures~7 shows that the high density cores of dark matter halos in
the equilibrium halo velocity dispersions are 10-20\% below the dark
matter's velocity dispersion. Figure~8 shows the half mass radii of
the tracers are reduced by a factor up to 3, and the virial radii are
reduced by a factor up to 5, below the dark matter values.  Note that
the ``galaxies'' have a rising relative $r_{1/2}$ with time, and,
larger $r_{1/2}$ than any other population--an indication that there
these objects are eroding in the central region of the cluster. There
is only the single central galaxy (a ``cD'') inside 100\hkpc, and 3
somewhat disrupted appearing objects between 100 and 200\hkpc.

An estimate of \bvs\ can be made on the basis of the data displayed in
Figure~7. The galaxy tracers are the relatively durable structures
visible in Figure~3. Averaging the 3 data points after the last merger
gives $\bvs\simeq0.7$. Averaging these tracers over all times after
5~Gyr, including the merger, gives a similar value. The average
velocity bias of the ``galaxies'' is $\bvs=0.85$. However this \bvs\
is an underestimate with a lower limit of $\bvs\ge0.75$ based on the
fraction of ``proto-galaxy'' particles unbound to ``galaxies'').  On
the basis of this and data from full cosmological simulations
(Couchman and Carlberg 1992) the best estimate of $\bvs=0.8\pm0.1$. A
simulation with $N\sim10^7$ in a single cluster should be able to
establish a conclusive value, within the context of dissipationless
clustering.

The two identifiable causes of the dynamical differences between the
tracer populations are a weak tendency for galaxies to form near the
cluster center, a manifestation of statistical bias (Kaiser 1984,
Bardeen \et 1986), and a dynamical bias loss of orbital energy
from the galaxy tracers.  The internal dispersions of
the condensed regions contribute about 1\% to \bvs, since the
typical galaxy halo has an internal velocity dispersion of about 100
\kms whereas the cluster has a dispersion of about 1000 \kms.
Velocity bias builds up during cluster
assembly between times 2.5~Gyr and 9.5~Gyr. Dense regions tend to
``sink'' relative to low density regions, leading to an energy
exchange (Quinn \et 1986), which is a form of dynamical friction in an
evolving potential.  When infall is smooth, it can be approximated as
energy conserving (Gott and Gunn 1972, Bertschinger 1985) whereas
infall of galaxies embedded in dark halos
having large internal density gradients leads
to a differential loss of energy with the dense cores becoming more
bound than the low density outer regions (West and Richstone 1988).

The radial distribution of the tracers is initially somewhat more
centrally concentrated than the total population (the initial points
on Figure~8).  Using the half mass radius to measure the statistical
formation bias (Kaiser 1984) finds that the luminosity weighted single
particle tracer populations are initially between 5 and 25\% more
compact in $r_{1/2}$ than the dark matter, but the numbers of galaxy
tracers are somewhat more extended than the dark matter
population. That is, big galaxies systematically form closer to the
center of the cluster, as is expected in statistical biasing.  As the
evolution of the cluster proceeds the small initial bias is completely
swamped by a dynamical bias which grows with time (Figure~8). The
individual particles become more concentrated to the center. Because
erosion is destroying ``galaxies'' near the center, the half mass
radius of the galaxies rises with time.  In spite of the erosion, the
virial radius of the ``galaxies'' always underestimates the true
virial radius. Thus the fact that the ``galaxies'' display a velocity
bias, even in this situation where numerical effects are working to
create an anti-bias, is strong evidence that velocity bias and mass
segregation will be even stronger in a larger N simulation, less
subject to galaxy evaporation. The individual particles, which trace
the light, remain a conservative estimator of the bias because
individual particles cannot merge to lower their translational
velocity, nor can they have an ``inverse slingshot'' transfer of
energy to a halo, which increases their binding energy to the cluster
(Heisler \& White 1990).

\setbox18=\vbox{
\noindent\ifapj\else\narrower\baselineskip 12pt\fi
Figure 9:\hskip 5mm A snapshot of all the particles in a
$0.2\overline{d}$ group at 14.4~Gyr, about one dynamical time after
infall. The box size is 4.6 \hmpc\ centered on the cluster center of
mass, and is comparable to Figure~1.  The core particles are the
tightly bound group at the lower left. The massive, low density halo
has been tidally removed and spread throughout the cluster.  }

\ifapj\else
\midinsert
\epsfysize=3.0truein
\centerline{\epsfbox[20 190 588 755]{strip.ps}}
\smallskip
\copy18 \endinsert\fi

\setbox19=\vbox{
\noindent\ifapj\else\narrower\baselineskip 12pt\fi
Figure 10:\hskip 5mm Evolution of the energy distribution for the
particles of Figure~9. Histograms of number of particles are plotted
against binding energy, measured in units of
$3.42\times10^4$~km$^2$s$^{-2}$.  The top panels are for the tightly
bound core, the lower panels are for all the particles, which are
dominated by the low density outer halo which is tidally removed from
the core.  The average binding energy of the halo particles does not
change significantly compared to the total mass distribution, however,
the average binding energy of the core particles increases about 22\%.
}
\ifapj\else
\midinsert
\epsfysize=4.0truein
\centerline{\epsfbox[24 200 588 688]{estrip.ps}}
\copy19 \endinsert \fi

\ls\ls\goodbreak
\ni\ub{\bf 5. COLLISIONLESS ENERGY TRANSFERS}

Galaxies orbiting in clusters are subject to dynamical friction which
slows heavy objects by heating the surrounding light
particles (Chandrasekhar 1942, Binney and Tremaine 1987).  Galaxies
fall into the cluster with massive extended halos, often bound in
small groups (see Figure~11).  Once in the cluster, tidal forces
unbind the galaxy groups and much of an individual
galaxy's outer dark halo.  The observation
that galaxies of different luminosities remain well mixed in a cluster
indicates that dynamical friction does not remain to any significant
degree once galaxies are orbiting inside the cluster (White 1976).
The evolution of the extended halos of galaxies are central to the
existence of a single galaxy velocity bias.

An xy snapshot of a group that was identified with $0.2\overline{d}$
at 2.5~Gyr is displayed at time 14.4~Gyr in Figure~9, and the
evolution of the particle energy distribution over that interval is
shown in Figure~10. The tight group is a ``galaxy'' of 82 particles
identified by joining particles closer than $0.2\overline{d}$ at
2.5~Gyr and then reselecting these particles closer than
$0.02\overline{d}$ at 13~Gyr. The core is a well organized
self-gravitating region, inside an outer halo of 714 particles
($1.6\times10^{12}h^{-1}\msun$) closer than $0.4\overline{d}$ at
2.5~Gyr. Linking at $0.47\overline{d}$ finds a massive structure
containing $4.4\times10^{14}h^{-1}\msun$ (192,000 particles) is linked
together.  Defining the mass of the halo as the number of particles in
a sphere which is moving no faster than half of the speed of the core
with respect to the cluster shows that up to time 9.4~Gyr the core has
an extended halo of $\gta 10^4$ particles.  The halo is truncated by
tidal forces to about 1500 particles at 9.4~Gyr, ending with about 400
particles around the central object at time 20~Gyr (Figure~11).  The
ratio of the time scale for dynamical friction to the crossing time is
$t_f/t_c \simeq 10^{-1} M_c/M_s$, where $M_c$ is the mass of the
cluster and $M_s$ is the mass of the satellite. For a cluster 100
times more massive than the infalling substructure the effect of
friction should be to remove of order 10\% of its binding energy,
which at this level of approximation is born out by the measurements
presented in Figure~10. The binding energy of [``halo'', ``core'']
particles is [-3.52, -3.48 ] ($\times10^6$ km$^2$s$^{-2}$) at 2.5~Gyr
and is [-4.61, -5.53] ($\times10^6$ km$^2$s$^{-2}$) at 19.7~Gyr (there
is little change after the first cluster crossing, at 10~Gyr).  The
binding energy of individual particles is not conserved in a time
varying Hamiltonian, but there is about a 22\% increase in the binding
energy of core particles in relation to the halo particles.

There are two possibilities for the origin of energy loss from the
core: energy exchange with the tidally stripped envelope (Heisler and
White 1990), which is independent of mass for a fixed structure, and,
friction on the halo which has an increasing effect with mass.  For
bulk of the mass, infall can be adequately approximated as
dissipationless, but the small fraction arriving in dense central
cores suffers a significant orbital energy transfer (Figure~10) to the
halo.  For the particles shown in Figures~9-11 it is not possible to
conclusively distinguish between the two effects, however the orbit is
not nearly as radial as Heisler and White (1990) studied, and some
work in progress on highly symmetrical collapses shows a strong
correlation of core energy loss with mass, pointing to the likely
dominance of friction in the core energy loss.

\setbox20=\vbox{
\noindent\ifapj\else\narrower\baselineskip 12pt\fi
Figure 11:\hskip 5mm The effective total mass
of a ``galaxy'' at different times.
The radial velocity dispersion of particles in
surrounding spheres is plotted  as a function of
mass enclosed centered on the 82 particle bound group used
in Figures~9 and 10.  At times up to 7.3~Gyr the surroundings
are moving coherently with the group. After 9.4~Gyr the correlated
halo is stripped off and the low velocity dispersion core orbits
through the hot background with little friction.}

\ifapj\else
\midinsert
\epsfysize=3.0truein
\centerline{\epsfbox[24 200 588 688]{mrsig.ps}}
\copy20 \endinsert \fi

\ls\ls\goodbreak\ni\ub{\bf 6. DISCUSSION AND CONCLUSIONS}

\ls
The main result of this paper is that if galaxies form at the centers
of dark matter halos via hierarchical gravitational instability then
current simulations are adequate to demonstrate that there must be a
single particle velocity bias, $b_v(1)\simeq0.80\pm0.1$, in rich,
reasonably relaxed, clusters.  This is most clearly demonstrated using
single particles as galaxy tracers, since simulated galaxies near the
cluster center are systematically eroded by two-body heating. The
velocity cooling leads to a shrinking of the galaxies' light
distribution with respect to the cluster dark matter, and virial
theorem estimators can report cluster masses nearly an order of
magnitude lower than the virialized mass present.  In the simulation
the most durable galaxy tracers give up to a factor of 5 underestimate
of the virial mass, with evaporation of galaxy tracers in the inner
0.5\hmpc\ limiting the size of the reduction.

Single particle velocity bias appears during infall into the
cluster. That is, there is a small statistical bias ($\sim$10\% in radius) of
galaxy tracers to the mass in the initial
conditions, after the first cluster crossing the dense central cores
of infalling halos have become systematically more bound to the
cluster than the low density outer halos.  With velocity bias, the
Cosmological Virial Theorem is adjusted to $\Omega b^2_v(2)/b$ (Fisher
\et 1993) which is measured to be 0.36 (IRAS galaxies, Fisher \et
1993) and 0.2 (optical, Davis \& Peebles 1983, Bean \et 1983). A
simple model of cluster anti-bias predicts that $\bvp\simeq\bvs-0.15$.
It is therefore possible to accommodate $\Omega=1$ if $\bvp\simeq0.6$
and $b\simeq1.0$ ($b=1/\sigma_8$) for IRAS galaxies and $b\simeq1.8$
for optical galaxies. Until the existence of velocity bias is
observationally established, the suggestion that $\Omega=1$ on the
basis of clustering dynamics alone remains insecure.  A testable
prediction of the single particle form of velocity bias is that
clusters should have extended massive halos, typically 2 to 3 times
the virial mass at 3 virial radii, which can be detected with
measurements of cluster gravity using velocities and weak
gravitational lensing.  The value of velocity bias required to reduce
the pairwise dispersions of a COBE normalized CDM spectrum (Smoot \et
1992, Wright \et 1992, Couchman and Carlberg 1992) which are about
1500 \kms\ in the mass field at 1 \hmpc, is $\bvp\simeq0.2$, well out
of the acceptable range found in this paper.

At the present time, there is no conclusive observational evidence for
the existence of any form of velocity bias, nor is it ruled out.  The
single particle velocity bias should exist only in virialized
regions. The primary prediction is that clusters of galaxies should
have extended dark matter halos, so that the cluster mass to light
ratio rises with radius (Figure~1).  Either infalling or virialized
galaxies should give an accurate estimate of the local potential
gradient (Villumsen and Davis 1986, Carlberg 1991) and hence the true
cluster mass inside their orbits. Weak gravitational lensing (\eg
Tyson \et 1990, Fahlman \et 1994, Bonnet \et 1994) is another powerful
tool. Measurement of a rise in mass-to-light is a difficult
observation which is confused by foreground and background objects,
triaxial cluster geometry, and aspherical surrounding large scale
structure. The main numerical uncertainty in the simulation presented
here is two-body heating which dissolves substructures in the core of
the cluster, which could be effectively eliminated in a simulation
having $10^7$ particles.

\ifapj \vfill\eject \fi
\goodbreak
\ls\ls
\ni\ub{REFERENCES}
\parskip=0pt

\hi{Bardeen, J. M., Steinhardt, P. J. \& Turner, M. S. 1983, Phys. Rev. D., 28,
679}
\hi{Bardeen, J. M., Bond, J. R., Kaiser, N. \& Szalay, A. S. 1986, \apj 304,
15}
\hi{Barnes, J., \& Hut, P. 1986, Nature 324, 446}
\hi{Bean, A. J., Efstathiou, G., Ellis, R. S., Peterson, B. A. \& Shanks, T.
1983, \mn 205, 605}
\hi{Bertschinger, E. 1985, \apjs 58, 39}
\hi{Bertschinger, E., Dekel, A., Faber, S. M., Dressler, A. \& Burstein, D.
1990, \apj  364, 370}
\hi{Bertschinger, E. \& Gelb, J. M. 1991, Computers in Physics, 5, 164}
\hi{Binney, J., \& Tremaine, S. 1987, {\it Galactic Dynamics}, (Princeton
University Press: Princeton)}
\hi{Blumenthal, G. R., Faber, S. M., Primack, J. R., \& Rees, M. J. 1984,
Nature 311, 517}
\hi{Bond, J. R., \& Efstathiou, G. E. 1984, \apjl 285, L45}
\hi{Bonnet, H., Mellier, Y., \& Fort, B. 1994, preprint}
\hi{Carlberg, R. G. 1991, \apj 367, 385}
\hi{Carlberg, R. G. \& Dubinski, J. 1991, \apj 369, 13}
\hi{Carlberg, R. G. \& Couchman, H. M. P. 1989, \apj 340, 47}
\hi{Carlberg, R. G., Couchman, H. M. P. \& Thomas, P. A. 1990, \apjl 352, L29}
\hi{Cen, R. \& Ostriker, J. P. 1992, \apjl 399, L113}
\hi{Chandrasekhar, S. 1942, {\it Principles of Stellar Dynamics}, (University
of Chicago Press: Chicago)}
\hi{Chernoff, D. and Weinberg, M. 1990, \apj 351, 121}
\hi{Couchman, H. M. P. \& Carlberg, R. G. 1992, ApJ 389, 453}
\hi{David, L, \& Blumenthal, G. 1992, \apj 389, 510}
\hi{Davis, M. \& Peebles, P. J. E. 1983, \apj {267}, 465}
\hi{Davis, M., Efstathiou, G., Frenk, C. S. \& White, S. D. M. 1985, \apj 292,
371}
\hi{Dubinski, J. University of Toronto M.~Sc. thesis, 1988}
\hi{Dubinski, J. \& Carlberg, R. G. 1991, \apj 378, 496}
\hi{Evrard, A. E., Summers, F. J. \& David, M. 1992, preprint}
\hi{Fahlman, G. G., Kaiser, N., Squires, G., \& Woods, D. 1994,
	\apjl submitted}
\hi{Fall, S. M. \& Rees, M. J. 1977, \mn 181, 37P}
\hi{Fisher, K. B., Davis, M., Strauss, M. A., Yahil, A., \& Huchra, J. P.
	1993, preprint}
\hi{Gelb, J. M. \& Bertschinger, E. 1993, Fermilab preprint}
\hi{Gott, J. R. \& Gunn, J. 1972, \apj 176, 1}
\hi{Gott, J. R., Turner, E. L., \& Aarseth, S. J. 1979, \apj 234, 13}
\hi{Guth, A. 1981, Phys. Rev. D., 23, 347}
\hi{Hale-Sutton, D., Fong, R., Metcalfe, N. and Shanks, T. 1989, \mn 237, 569}
\hi{Heisler, J. \& White, S. D. M. 1990, \mn  243, 199}
\hi{Hernquist, L. 1990, \apj 356, 359}
\hi{Hernquist, L. 1992, \apj 400, 460}
\hi{Huang, S., Dubinski, J, \& Carlberg, R.~G. 1993, \apj 404, 74}
\hi{Kaiser, N. 1984, \apjl 284, L9}
\hi{Kaiser, N., Efstathiou, G., Ellis, R., Frenk, C., Lawrence, A.,
Rowan-Robinson, M., \& Saunders, W. 1991, \mn 252, 1}
\hi{van Kampen, E. 1993, preprint}
\hi{Katz, N., Hernquist, L., \& Weinberg, D. 1992, \apjl 399, L109}
\hi{Katz, N. \& White, S. D. M. 1993, \apj 412, 455}
\hi{Kent, S. \& Gunn, J. E., 1982, \aj 87, 945}
\hi{Lynden-Bell, D., Lahav, O., \& Burstein, D. 1989, \mn 241, 235}
\hi{Nusser, A. \& Dekel, A. 1993, \apj 405, 437}
\hi{Peebles, P. J. E. 1976, \apjl 205, L109}
\hi{Peebles, P. J. E. 1980, {\it The Large-Scale Structure of the
Universe}, Princeton University Press}
\hi{Press, W. H. \& Schechter, P. 1974, \apj {\bf 187}, 425}
\hi{Quinn, P. J., Salmon, J. K., and Zurek, W. H. 1986, \apj 322, 392}
\hi{Rees, M. J. and Ostriker, J. P. 1977, \mn 179, 541}
\hi{Smoot, G. F., Bennett, C. L., Kogut, A., Wright, E. L., Aymon, J.,
Boggess, N. W., Cheng, E. S., De Amici, G., Gulkis, S., Hauser, M. G.,
Hinshaw, G., Jackson, P. D., Jannsen, M., Kaita, E., Kelsall, T.,
Keegstra, P., Lineweaver, C., Lo wenstein, K., Lubin, P., Mather, J.,
Meyer, S. S., Moseley, S. H., Murdock, T., Rokke, L., Silverberg,
R. F., Tenorio, L., Weiss, R., \& Wilkinson, D. T.  1992, \apjl 326,
L1}
\hi{Spitzer, L. 1962, {\it Physics of Fully Ionized Gases}, 2nd ed.
(Interscience: New York)}
\hi{Strauss, M. A., Yahil, A., Davis, M., Huchra, J. P., \& Fisher, K. 1992,
	\apj 397, 395}
\hi{Summers, F. 1993 Ph.~D. Thesis, University of California at Berkeley}
\hi{Tyson, J. A., Valdes, F., \& Wenk, R. A. 1990, \apjl 349, L1}
\hi{Villumsen, J. V. \& Davis, M. 1986, \apj 308, 743}
\hi{West, M. J., \& Richstone, D. O. 1988, \apj 335, 532}
\hi{White, S. D. M. 1976, \mn 177, 717}
\hi{White, S. D. M., Huchra, J. P., Latham, D., \& Davis, M. 1983,
	\mn 203, 701}
\hi{White, S. D. M. and Rees, M. J. 1978, \mn 183, 341}
\hi{White, S. D. M., Davis, M., Efstathiou, G. \& Frenk, C. S. 1987, Nature,
330, 451
\hi{Wright, E. L., Meyer, S. S., Bennett, C. L., Boggess, N. W.,
Cheng, E. S., Hauser, M. G., Kogut, A., Lineweaver, C., Mather, J. C.,
Smoot, G. F., Weiss, R., Gulkis, S., Hinshaw, G., Jannssen, M.,
Kelsall, T., Lubin, P. M., Moseley, S. H . Jr., Murdock, T. L.,
Shafer, R. A., Silverberg, R. F., \& Wilkinson, D. T. 199 2, \apjl
396, L13.}  }

\ifapj

\vfill \eject
\medskip
\copy22
\bigskip \bigskip \bigskip
\centerline{FIGURE CAPTIONS}
\medskip
\copy17
\bigskip
\copy21
\bigskip
\copy11
\bigskip
\copy12
\bigskip
\copy13
\bigskip
\copy14
\bigskip
\copy15
\bigskip
\copy16
\bigskip
\copy18
\bigskip
\copy19
\bigskip
\copy20

\vfill \eject
\centerline{\caps Fig. 1}
\epsfysize=7.5truein
\centerline{\epsffile{den.PS}}
\vfill\eject
\centerline{\caps Fig. 2}
\epsfysize=7.5truein
\centerline{\epsfbox[310 215 570 780]{one.PS}}
\vfill\eject
\centerline{\caps Fig. 3}
\bigskip
\epsfysize=7.5truein
\centerline{\epsffile{pdm.384.ps}}
\vfill\eject
\centerline{\caps Fig. 4}
\epsfysize=8.5truein
\centerline{\epsffile{pdsc.384.ps}}
\vfill\eject
\epsfysize=7.5truein
\centerline{\caps Fig. 5}
\centerline{\epsffile{vm.PS}}
\vfill \eject
\centerline{\caps Fig. 6}
\epsfysize=7.5truein
\centerline{\epsffile{vmb.PS}}
\vfill \eject
\centerline{\caps Fig. 7}
\epsfysize=7.5truein
\centerline{\epsffile{vvb.PS}}
\vfill \eject
\centerline{\caps Fig. 8}
\epsfysize=7.5truein
\centerline{\epsffile{vrb.PS}}
\vfill \eject
\centerline{\caps Fig. 9}
\epsfysize=7.5truein
\centerline{\epsffile{strip.PS}}
\vfill\eject
\centerline{\caps Fig. 10}
\epsfysize=7.5truein
\centerline{\epsffile{estrip.PS}}
\vfill\eject
\centerline{\caps Fig. 11}
\epsfysize=7.5truein
\centerline{\epsffile{mrsig.PS}}
\fi

\bye